\documentclass[aps,prb,superscriptaddress,twocolumn,nofootinbib]{revtex4-2}
\usepackage[pdftex]{graphicx}
\usepackage{amsfonts}
\usepackage{amsthm}
\usepackage{amsmath}
\usepackage{amsfonts}
\usepackage{amssymb}
\usepackage{bm}
\usepackage{braket}
\usepackage[normalem]{ulem}
\usepackage[dvipsnames]{xcolor}
\usepackage[colorlinks=True,linkcolor=OrangeRed,citecolor=BlueViolet,urlcolor=BlueViolet]{hyperref}

\usepackage{float}

\newcommand\new[1]{\black{#1}}
\newcommand\eq[1]{\begin{align}#1\end{align}}

\newcommand\black[1]{{\color{black}{#1}}}

\newcommand{\deltat}{\Delta_\mathrm{typ}}
\newcommand{\yt}{y_\mathrm{typ}}
\newcommand{\w}{\omega}
\newcommand{\me}{\omega_\mathrm{ME}}
\newcommand{\pd}{\phantom\dagger}
\newcommand{\comment}[1]{}

\begin{document}
\title{Localisation in quasiperiodic chains:\\ 
a theory based on convergence of local propagators}
\author{Alexander Duthie}
\email{alexander.duthie@chem.ox.ac.uk}
\affiliation{Physical and Theoretical Chemistry, Oxford University,
South Parks Road, Oxford OX1 3QZ, United Kingdom}

\author{Sthitadhi Roy}
\email{sthitadhi.roy@chem.ox.ac.uk}
\affiliation{Physical and Theoretical Chemistry, Oxford University,
South Parks Road, Oxford OX1 3QZ, United Kingdom}
\affiliation{Rudolf Peierls Centre for Theoretical Physics, Clarendon Laboratory, Oxford University, Parks Road, Oxford OX1 3PU, United Kingdom}

\author{David E. Logan}
\email{david.logan@chem.ox.ac.uk}
\affiliation{Physical and Theoretical Chemistry, Oxford University,
South Parks Road, Oxford OX1 3QZ, United Kingdom}
\affiliation{Department of Physics, Indian Institute of Science, Bangalore 560012, India}

\begin{abstract}
Quasiperiodic systems serve as fertile ground for studying localisation, due to their propensity already in one dimension to
exhibit rich phase diagrams with mobility edges. The deterministic and strongly-correlated nature of the quasiperiodic potential 
nevertheless offers challenges distinct from disordered systems. Motivated by this, we present a theory of localisation in quasiperiodic chains with nearest-neighbour hoppings, based on the convergence of local propagators; exploiting the fact that the imaginary part of the  associated self-energy acts as a probabilistic order parameter for localisation transitions and, 
importantly, admits a continued-fraction representation. Analysing the convergence of these continued fractions, 
localisation or its absence can be determined, yielding in turn the critical points and mobility edges. Interestingly, we find anomalous scalings of the order parameter with system size at the critical points, consistent with the fractal character of critical eigenstates. 
Self-consistent theories at high orders are also considered, shown to be conceptually connected to the theory based on continued fractions,
and found in practice to converge to the same result. 
Results are  exemplified by analysing the theory for three  quasiperiodic models covering a range of behaviour.
\end{abstract}
\maketitle


\section{Introduction\label{sec:intro}}

The physics of localisation in disordered quantum systems~\cite{anderson1958absence} has been a cornerstone of condensed matter 
theory and statistical mechanics  for well over half a century. It is however well understood by now  that the paradigm of localisation goes beyond systems with quenched random disorder: systems with \emph{quasiperiodicity} comprise a family of non-random and deterministic systems which host localisation and related phenomena such as mobility edges, robust multifractality, ‘mixed phases’ with both extended and localised eigenstates, and anomalous transport~\cite{aubry1980analyticity,Harper_1955,sokoloff1980electron,prange1983wave,Kohmoto1983metalinsulator,sarma1988mobility,boers2007mobility,biddle2009localization,biddle2010predicted,biddle2011localization,ganeshan2015nearest,Morales2014resonant,Lahini2009observation,Modak2016integrals,wang2017quantumwalks,li2017mobility,gopalakrishnan2017self,Purkayastha2017nonequilibrium,Purkayastha2018anomalous,roy2018multifractality,luschen2018single,YaoPRL2019,sutradhar2019transport,wang2020onedimensional,wang2020duality}, as well as many-body localised phases in the presence of interactions~\cite{iyer2013manybody,li2015manybody,modak2015manybody,khemani2017two}.

In phenomenological terms, localisation in quasiperiodic systems is quite different from, and arguably richer than, that occurring in their disordered cousins.   For instance, the simplest quasiperiodic model, the Aubry-Andr\'e-Harper (AAH) model~\cite{aubry1980analyticity,Harper_1955}, hosts a localisation transition already in one-dimension, and variants of 
the model have genuine mobility edges in their spectra~\cite{prange1983wave,sarma1988mobility,boers2007mobility,biddle2009localization,biddle2010predicted,biddle2011localization,ganeshan2015nearest,li2017mobility,YaoPRL2019,wang2020onedimensional,wang2020duality}. In fact, mobility edges appear quite typically in systems where the quasiperiodicity arises from a continuous periodic potential incommensurate with the underlying periodic lattice.\footnote{There are of course other quasiperiodic models which are critical throughout their spectra and parameter space, such as Fibonacci chains~\cite{kohmoto1983localization,ostlund1983one}.}
In this regard the AAH model itself is a special case in which,  due to an exact energy-independent duality,
eigenstates at all energies undergo a localisation transition at the same point, such that there is no genuine mobility edge.

From a theoretical point of view, quasiperiodic systems are also qualitatively different to disordered ones, because the
potential in the former is deterministic and hence infinite-range correlated. Much of the remarkable theoretical progress in disordered systems over the years, stems from the ability to average over uncorrelated disorder in an independent 
and unbiassed fashion. In this regard, quasiperiodic systems pose a unique challenge: the deterministic nature of the potential implies the need to account for the potential at all points in space simultaneously. One thus expects the analysis involved to be bespoke to the specific model considered. This is indeed typically the case; examples include model-specific energy-dependent generalised duality transformations~\cite{biddle2010predicted,biddle2011localization,ganeshan2015nearest}, or duality transformations relating models with known phase diagrams~\cite{wang2020duality}, and Lyapunov exponent calculations based on global theories of Schr\"odinger operators~\cite{wang2020onedimensional}.

It is therefore of importance to develop a general theoretical framework to predict and analyse the localisation phase diagrams, 
\new{as well as to understand theoretically the nature of the phases therein,}
for essentially arbitrary quasiperiodic models. A step towards that was taken by us in a recent 
work~\cite{duthie2020selfconsistent}, where a \new{leading-order} self-consistent theory for mobility edges was developed. The theory, of a self-consistent mean-field nature, was rooted in analysis of the local propagators and in particular the imaginary part of their self-energies, using a renormalised perturbation series (RPS)~\cite{feenberg1948note,Economoubook,abou-chacra1973self} at leading order. In spirit, the theory was inspired heavily from  self-consistent approaches to localisation in disordered systems~\cite{abou-chacra1973self,economous1972existence,logan1987dephasing}. \new{The leading-order self-consistent theory gave analytical access to the mobility edges and critical points for essentially arbitrary quasiperiodic models. At the same time, it raised a number of significant
conceptual questions:
\begin{enumerate}
\item Is the leading-order theory robust to the addition of higher-order terms?
\item Can higher-order mean-field theories recover what is missed at first-order, in particular the presence of hierarchical spectral gaps and 
related 
finer structures in the distributions of self-energies?
\item What is the fate of the mean-field theories if taken to infinite order?
\end{enumerate} }
\new{Answering these questions necessarily requires us to go beyond the leading-order treatment of Ref.~\cite{duthie2020selfconsistent}, and construct a framework for higher-order theories which takes into account the correlated and deterministic potential at all sites simultaneously. This constitutes the central motivation of our work.
}

Exploiting the fact that the RPS for the \new{local} self-energies can be recast as a continued fraction  (CF), we analyse the latter's convergence \new{-- a treatement that is close in spirit to that of  Anderson's original work~\cite{anderson1958absence} on localisation. The CF at any arbitrary order $\ell$ explicitly takes as input the quasiperiodic potentials at all sites within a distance $\ell$ from the site in question.} 
Additionally, we also truncate the continued fraction at arbitrary order, and perform a self-consistent analysis at that order; which yields quantitatively
the same results for the localisation phase diagram as the convergence of the continued fraction, as well as very
good agreement with the leading-order theory.
Going to higher orders 
uncovers finer structures within the phases which are not obtained at leading-order level, such as hierarchical spectral gaps and additional structures in the distributions of the self-energies, which reflect the deterministic, quasiperiodic nature of the potentials.
It also makes a case for the robustness of the leading-order theory with regard to the location of mobility edges.
We note here that the results obtained are found to be in excellent agreement with numerical results obtained from exact diagonalisation for a broad class of quasiperiodic chains.

We turn now to an overview of the paper.



\subsection*{Overview}

In order to test the theory, we employ three models with exactly known mobility edges. Defined and described briefly in  
Sec.~\ref{sec:models}, these are chosen to span a wide range of behaviour, from no mobility edge  (transition at same critical point for all energies) to multiple mobility edges in the spectra.

Section~\ref{sec:localprop} is devoted to setting up the basic formalism for the local propagator and the self-energies which 
underpin the work. In particular, we lay special emphasis on the imaginary part of the local self-energy $(\Delta(\omega))$, and
in Sec.~\ref{sec:imse} discuss its importance as a probabilistic order parameter for a localisation transition.  Physically, $\Delta(\omega)$ for any site is the loss rate (or inverse lifetime) of probability amplitude from that site, into  eigenstates of energy $\omega$ which overlap the site. It has long served as a  powerful tool for studying localisation and localisation transitions in disordered systems~\cite{abou-chacra1973self,economous1972existence}.

With unit probability, $\Delta(\omega)$ is respectively finite and vanishing in an extended and localised  
phase~\cite{abou-chacra1973self}.  Crucially, in the latter, it vanishes $\propto\eta\to 0^{+}$ (with $\eta$ the imaginary part of the energy);
this allows one to define a further probabilistic order parameter 
$y(\omega)=\Delta(\omega)/\eta$, which is finite in the localised phase and divergent in the extended.
Before delving into a detailed analysis, Sec.~\ref{sec:numres} gives numerical results for $\Delta(\omega)$ and $y(\omega)$ obtained via exact diagonalisation, as a demonstration of their validity as order parameters. In Sec.~\ref{sec:cf}, we discuss how the RPS for $\Delta(\omega)$ and $y(\omega)$ can be recast as a continued fraction, which forms the basis of the analysis in subsequent sections. 

Analysis of the convergence of the continued fractions constitutes Sec.~\ref{sec:convcf}. We show that in the localised phase, the continued fraction for  $y(\omega)$ converges and has a finite typical value, $\yt(\w)$ (the geometric mean of its distribution); whereas in the extended phase it does not, and the typical $y(\omega)$ is divergent in the thermodynamic limit. Using this diagnostic for the quasiperiodic chains considered, one can determine the presence of localised or extended states at any point in the parameter and energy space. This is the first main result of the work.

In Sec.~\ref{sec:sct} we present a self-consistent theory at arbitrary orders by truncating the continued fractions; 
which is the second main result of the work. Consistent with the results obtained in Sec.~\ref{sec:convcf}, the self-consistent typical $y(\omega)$ is respectively finite and divergent in the localised and extended phases. Equivalently, the typical 
$\Delta(\omega)$ obtained self-consistenly is finite and vanishing in the extended and localised phases respectively. 
An important point to note is that the convergence of the continued fraction for $y(\omega)$ is tied to the convergence of the self-consistent typical $\yt(\omega)$ to a finite value. The analysis presented places the recently investigated
leading-order theory~\cite{duthie2020selfconsistent} within a broader framework, encompassing  both  higher-order self-consistent theories as well as the convergence of the underlying RPS for $y(\omega)$.

\begin{figure}
\includegraphics[width=\linewidth]{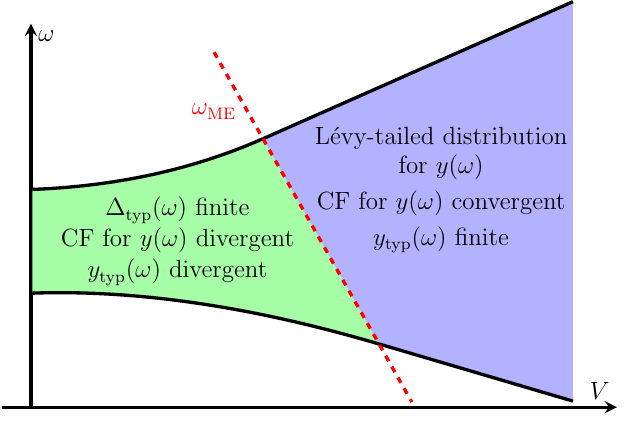}
\caption{Schematic localisation phase diagram in the space of energy ($\omega$) and quasiperiodic potential strength $(V)$. A mobility edge (red dashed line) separates the extended phase (green) from the localised (blue). The salient features of the imaginary part of the self-energy in the two phases ($\Delta(\w)$ and $y(\w)=\Delta(\w)/\eta$) are indicated. CF denotes continued fraction. }
\label{fig:schematic}
\end{figure}

As mentioned above, $\Delta(\omega)$ and $y(\omega)$ are probabilistic order parameters. Their distributions are thus of fundamental interest, and this forms the subject of Sec.~\ref{sec:dists}. We obtain the distributions using both the continued fractions as well as self-consistently; both of which show excellent mutual agreement, as well as with results obtained from exact diagonalisation. One main result  here is that the distributions of $y(\omega)$ in the localised phase have L\'evy tails 
$\propto y^{-3/2}$; which likewise arise in localised phases of disordered systems, for both 
uncorrelated~\cite{abou-chacra1973self} and correlated disorder~\cite{roy2020localisation}, suggesting they are rather universal in localised systems. It is also of course because of these fat-tails  that the geometric mean ($\yt$) is a suitable measure of typicality for the distribution.

We close in Sec.~\ref{sec:discussion} with some discussion and directions for future work.


\paragraph*{Summary of results:}

\new{The essential outcome of this work is a theory of localisation in quasiperiodic chains with nearest-neighbour hoppings, based on the convergence properties of the local propagators and associated self-energies. 
The extended and localised regimes can be diagnosed in terms of $\Delta(\omega)$ and $y(\omega)$ as follows. Where the CF for $y(\omega)$ converges, localised states are present and $\yt$, whether obtained from the CF or self-consistently, takes a finite value. A physical insight here is that once the order of CF or the self-consistent theory is much larger than the localisation length, the two approaches necessarily converge to the same result for $\yt$, which in turn gives the exact result. $\deltat$ is naturally vanishing 
for a localised phase in the $(V,\omega)$-plane where eigenstates exist (with $V$ the quasiperiodic potential strength).
Conversely, a finite $\deltat$ in the thermodynamic limit
implies the presence of extended states and a concomitant divergence in the CF for $y(\omega)$, reflecting the lack of a self-consistent solution for $\yt(\omega)$. The critical points/mobility edges in the $(V,\omega)$-plane are therefore unambiguously identified as points where $\deltat$ vanishes and $\yt$ diverges simultaneously. A schematic summary is given in in Fig.~\ref{fig:schematic}, showing a localisation phase-diagram with a mobility edge, with the salient features of the two phases indicated. }


\section{Models \label{sec:models}}

We consider quasiperiodic chains of length $L$ with nearest-neighbour hoppings, described generally by a Hamiltonian of form
\eq{
    H=V\sum_{j=0}^{L-1}\epsilon_j^{\phantom\dagger} c_j^\dagger c_j^{\phantom\dagger} +
    J\sum_{j=0}^{L-2}[c_j^\dagger c_{j+1}^{\phantom\dagger}+ \mathrm{H.c.}] \,, 
    \label{eq:genham}
}
with $c_j^\dagger~(c_j)$ the creation (annihilation) operator on site $j$. The model-specific quasiperiodic potential is encoded in $\epsilon_j$, with $V$ its strength, and $J$ the hopping amplitude; without loss of generality we consider $V,J\geq 0$.
We focus on three models which between them display a wide range of behaviour and for which the mobility edges $\me(V)$, or equivalently the energy-dependent critical points $V_c(\omega)$, are exactly known.

\paragraph*{The Aubry-Andr{\'e}-Harper model:} The first is the familiar and much-studied AAH 
model~\cite{aubry1980analyticity,Harper_1955} defined via 
\begin{equation}
    \epsilon_j^{\phantom\dagger} = \cos(2\pi\kappa j+\phi) \, , 
    \label{eq:AAH-potential}
\end{equation}
with $\kappa$ an irrational number, reflecting the incommensurability of the potential relative to the underlying lattice
(here we take $\kappa$  to be the golden mean); and $\phi \in [0,2\pi)$ is a random but uniform phase shift employed to perform the analogue of disorder averaging. The model does not host a genuine mobility edge, as all eigenstates undergo a localisation transition at the critical  $V_c = 2J$ (localised for $V>V_c$)~\cite{aubry1980analyticity,Harper_1955}. Equivalently, the mobility edge can be viewed as a line parallel to the energy axis at $V=V_c$.

\paragraph*{The $\beta$-model:} The second model was introduced in Ref.~\cite{ganeshan2015nearest}, and we refer to it as the $\beta$-model. It is a variant of the AAH model, and is described by
\eq{
\epsilon_j^{\pd} = \frac{\cos(2\pi\kappa j+\phi)}{1-\beta\cos(2\pi\kappa j+\phi)}\,,
\label{eq:beta-potential}
}
with $0\le\vert\beta\vert<1$. The potential in Eq.~\eqref{eq:beta-potential} breaks the exact duality of the AAH model leading to a genuine mobility edge given by~\cite{ganeshan2015nearest}
\begin{equation}
\omega_\text{ME}^{\pd} = (2J-V)/\beta\,.\label{eq:beta_mobility_edge}
\end{equation}
In the limit $\beta\to 0$, the potential in Eq.~\eqref{eq:beta-potential} reduces to the AAH potential Eq.~\eqref{eq:AAH-potential} and, consistently, the mobility edge in Eq.~\eqref{eq:beta_mobility_edge} becomes a straight line parallel to the $\omega$-axis at $V=2J$.

\paragraph*{The mosaic models:} The third model is from the family of so-called mosaic models~\cite{wang2020onedimensional}, the name arising from the fact that regularly spaced blocks of sites have $\epsilon_j=0$ while the complementary sites experience a quasiperiodic potential. Explicitly,
\eq{
    \epsilon_j^{\pd} = \begin{cases}
                    \cos(2\pi\kappa j +\phi)~~~:~j=lk \\
                    0~~~~~~~~~~~~~~~~~~~~:~\mathrm{otherwise}
                \end{cases}
    \label{eq:mosaic-model}
}
where $k\in\mathbb{Z}$.  The system thus consists of blocks of size $l$ within which only the first site has a non-zero (AAH) potential. Throughout the rest of the paper we will focus on $l=2$, for which the system has two symmetric mobility edges~\cite{wang2020onedimensional}
\begin{equation}
\omega_\text{ME}^{\pd}=\pm{2J^{2}/V}\,.\label{mosaic_mobility_edge}
\end{equation}
Note from Eq.~\eqref{mosaic_mobility_edge} that states in the centre of the spectrum ($\omega=0$) are extended for all finite values of $V$.


\section{Local propagators and self-energies \label{sec:localprop}}
In this section  we set up the basic formalism for the local propagator and associated self-energies, which are the central quantities of interest in our theory. 


\subsection{Imaginary part of local self-energy \label{sec:imse}}

In the time domain, the local propagator at site $j$ is defined as $G_j(t)=-i\Theta(t)\braket{j|e^{-iHt}|j}$, which is 
simply the return probability amplitude. Since a localisation transition is an eigenstate phase transition -- and 
the very notion of a mobility edge is intrinsically energy dependent -- it is more natural to consider the local propagator in the energy domain, 
\begin{equation}
 G_j^{\pd}(\omega) = [\omega^+-V\epsilon_j^{\pd}-S_j^{\pd}(\omega)]^{-1},\label{eq:S_j_def}
 \end{equation} 
 where $\omega^+ = \omega+i\eta$ with $\eta=0^+$ the regulator, and $S_j(\omega)$ is the local self-energy of site $j$. 
 The self-energy has real and imaginary parts
 \begin{equation}
 S_j^{\pd}(\omega) = X_j^{\pd}(\omega) - i\Delta_j^{\pd}(\omega),
 \end{equation}
 where the real part can be interpreted physically as the perturbative shift in the eigenvalues from the bare $\epsilon_j$'s.
In physical terms the imaginary part, $\Delta_j(\omega)$,  gives the rate of loss of probability amplitude from site $j$ into eigenstates of energy $\w$ which overlap that site. It thus acts as a probabilistic diagnostic for the localisation transition, and our analysis centres on it. As mentioned previously, an extended (localised) phase is signified by a finite (vanishing) $\Delta_j(\omega)$ with unit probability. In particular, in the localised phase $\Delta_j(\omega)\propto \eta$, and as such one can define an equivalent probabilistic order parameter $y_j(\omega)=\Delta_j(\omega)/\eta$, which is finite with unit probability in the localised phase and diverges as the transition is approached from the localised side.

 The RPS  for $S_j(\omega)$~\cite{feenberg1948note,abou-chacra1973self,Economoubook} is given for a chain with nearest neighbour hoppings  by $S_j(\omega)=\sum_{k}J_{jk}^2G_k^{(j)}(\omega)$, with $J_{jk}$ the hopping amplitude between sites $j$ and $k$, and $G_k^{(j)}(\omega)$ the local propagator for site $k$ with site $j$ removed. Since any site $j$ only has two neighbours, 
$j\pm 1$, connected by a hopping $J$, $S_j(\omega)$ thus reads
\begin{equation}
    S_j^{\pd}(\omega)=J^2[G_{j-1}^{(j)}(\omega) + G_{j+1}^{(j)}(\omega)]~.
    \label{eq:Sj-rps-chain}
\end{equation}
Quite crucially, $G_{j\pm 1}^{(j)}$ are the local propagators of sites $j\pm1$ with the site $j$ removed, and hence depend only on the sites to the right/left of $j$. They are  the \emph{end-site} propagators of two semi-infinite \textit{half-chains}. The decomposition in Eq.~\ref{eq:Sj-rps-chain} thus allows $S_j(\omega)$ to be constructed entirely from the knowledge of the half-chain propagators alone, which will be the focus of our analysis henceforth.

To simplify notation, in the following  we will denote the end-site propagator as $G_0(\omega)$ and the associated self-energy as $S_0(\omega)$. Since the chain end-site, $j=0$,   is connected to only one site, $j=1$, the RPS for $S_0(\omega)$ has only one term, $S_0(\omega) = J^2G_1^{(0)}(\omega)$ where $G_1^{(0)}(\omega)$ is again the end-site propagator of a chain, now starting at $j=1$. This recursive structure allows us to write $G_0(\omega)$ as a continued fraction
\begin{equation}
G_{0}^{\pd}(\omega) = \cfrac{1}{\omega^+-V\epsilon_{0}^{\pd}-\cfrac{J^2}{\omega^+-V\epsilon_{1}^{\pd}-\cfrac{J^2}{\ddots}}}\,,
\label{eq:Gcontf}
\end{equation}
from which the continued fraction for $S_0(\omega)$ follows as
\eq{
    S_{0}^{\pd}(\omega) = \cfrac{J^2}{\omega^+-V\epsilon_{1}^{\pd}-\cfrac{J^2}{\omega^+-V\epsilon_{2}^{\pd}-\cfrac{J^2}{\ddots}}}\,.
\label{eq:Scontf}
}
Equations \eqref{eq:Gcontf} and \eqref{eq:Scontf} are formally exact and, importantly, take into account the site energies at all sites $j$, crucial for the deterministic and infinite-range correlated quasiperiodic potentials.
We will turn to a detailed analysis of the continued fractions in Secs.~\ref{sec:cf} and \ref{sec:convcf}.

It is important to  emphasise that $\Delta_0(\omega)$ and $y_0(\omega)$ are probabilistic order parameters, as they are
characterised by distributions over values of $\phi$ and \emph{kinds} of end sites;\footnote{We refer to a model as possessing multiple \emph{kinds} of end sites if, depending on the configuration of the chain, the end sites experience different potentials. For instance, the $l=2$ mosaic model has two kinds of end sites, one with zero potential, and one with an AAH potential.} and that it is their typical (and not average) values that are appropriate order parameters. These can be obtained from their distributions $P_\Delta(\Delta_0)$ as
\eq{
    \Delta_\mathrm{typ}(\omega) = \exp\left[\int d\Delta_0^{\pd}~P_\Delta^{\pd}(\Delta_0^{\pd})\ln \Delta_0^{\pd}\right],
    \label{eq:deltatyp}
}
and similarly for $\yt(\omega)$.
Note further
that $\Delta_0(\omega)$ is proportional to the local density of states (LDoS), which is further suggestive that it is the typical value of $\Delta_0(\omega)$ which is an appropriate order parameter. Indeed, the typical LDoS has served as an order parameter for localisation transitions in disordered~\cite{logan1987dephasing,janssen1998statistics,dobrosavljevi2003typical} as well as quasiperiodic systems~\cite{ganeshan2015nearest}. The average value of $\Delta_0(\omega)$ by contrast is finite in both phases -- it merely gives the average density of states/eigenvalues -- and does not therefore discriminate between them.


\subsection{Results from exact diagonalisation \label{sec:numres}}

To demonstrate the validity of $\deltat(\omega)$ and $\yt(\omega)$ as suitable diagnostics of localised and extended phases, we present numerical results for them obtained via exact diagonalisation.  In terms of the half-chain eigenstates $\ket{\psi_n}$ and eigenvalues $E_n$, the local propagator $G_0(\omega)$ is 
\eq{
    G_0^{\pd}(\omega) = \sum_{n=1}^L\frac{\vert\braket{0|\psi_n}\vert^2}{\omega^+-E_n},
}
from which $S_0(\omega)$ can be trivially extracted using $G_{0}(\w) =[\w^{+}-V\epsilon_{0} -S_{0}(\w)]^{-1}$.
Since the regulator $\eta$ should be on the order of the mean level spacing, we take $\eta =c/L$ with $c\sim \mathcal{O}(1)$.

\begin{figure}
\includegraphics[width=\linewidth]{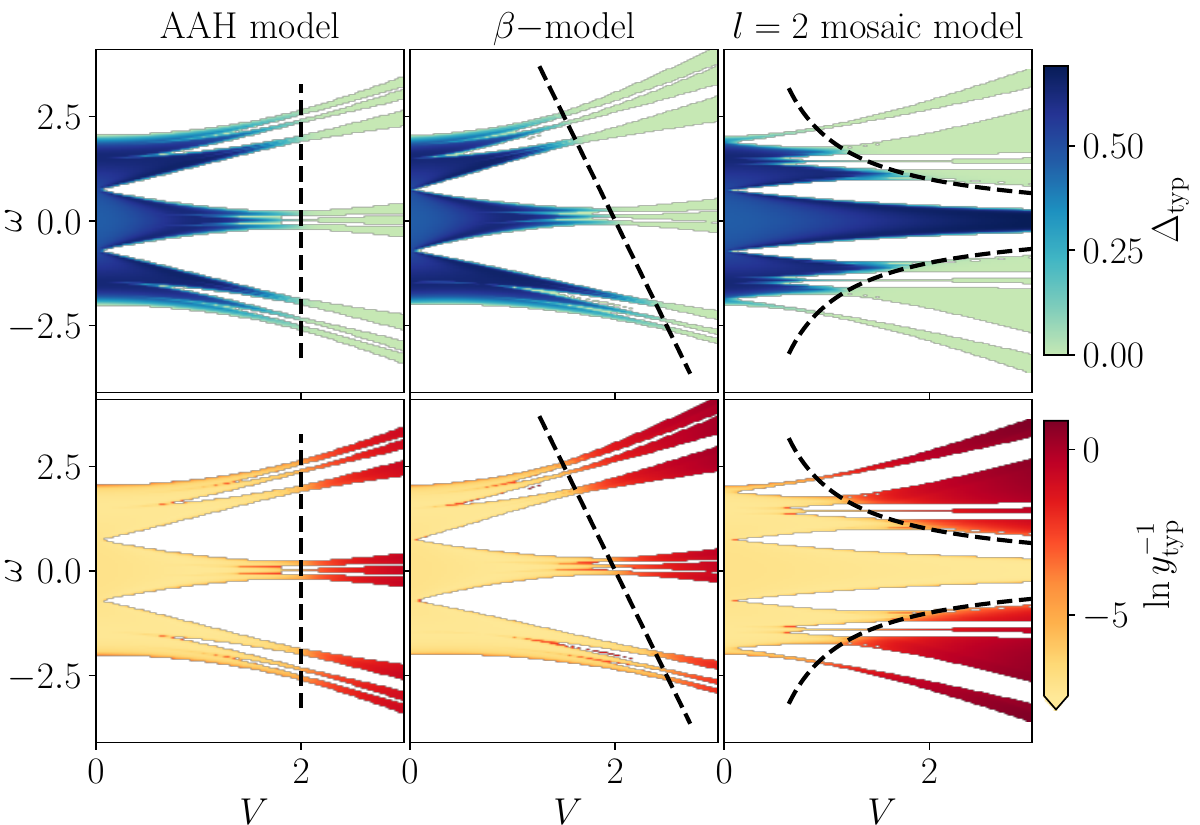}
\caption{Spectra obtained from exact diagonalisation for the three models described in Sec.~\ref{sec:models}, colour-coded with $\deltat(\omega)$ (top row) and $\yt(\omega)$ (bottom row), reflecting the localisation phase diagram and the mobility edges 
(black dashed lines). The columns correspond respectively to the AAH, the $\beta$ ($=0.2$), and the $l=2$ mosaic model.
A finite and vanishingly small value of $\deltat$ in the upper row indicates respectively extended and localised states. Consistently, $\yt$ is divergent and finite in the two phases as can be seen in the lower row. For clarity, note that we show $\ln \yt^{-1}$, which vanishes in the extended phase. All data shown for $L=2500$ with $\eta=1/L$, and statistics accumulated over 5000 values of $\phi\in[0,2\pi)$. Here, and in all subsequent figures, we set the hopping $J\equiv 1$. 
}
\label{fig:colourmap}
\end{figure}

\begin{figure}
\includegraphics[width=\linewidth]{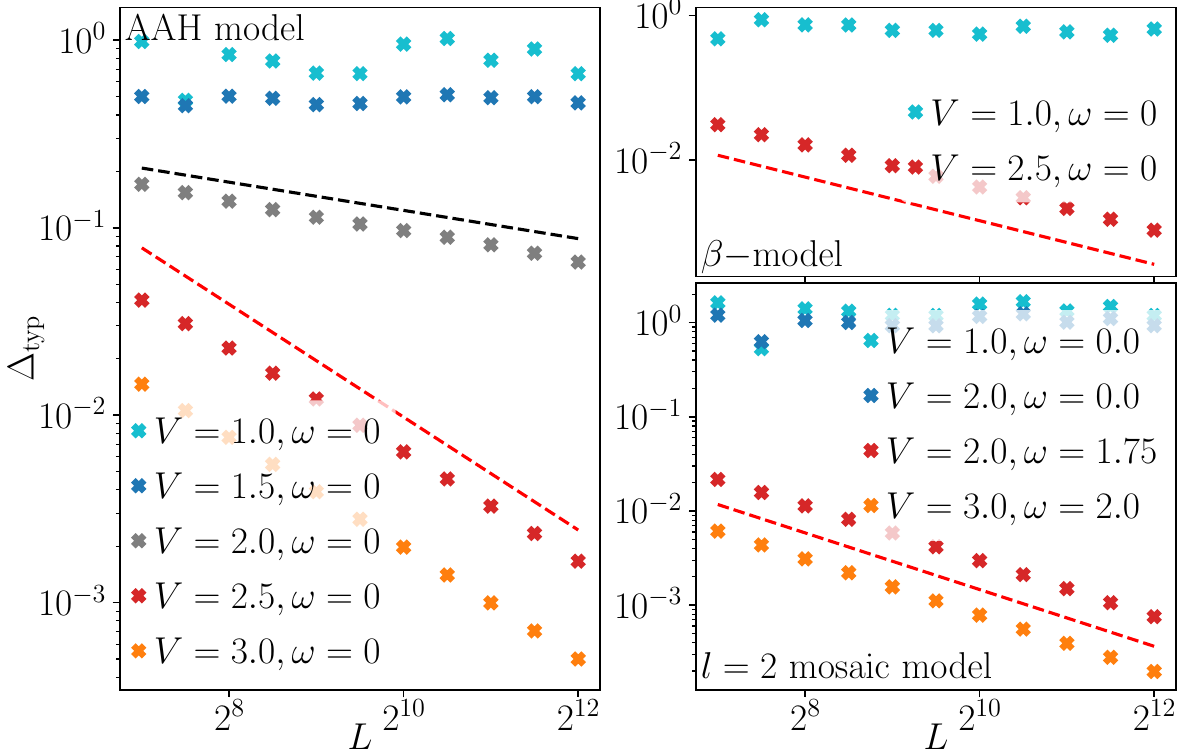}
\caption{System-size $(L)$ dependence of $\deltat(\omega)$ at representative $(V,\omega)$-points in the extended and localised regimes for all three models, obtained via exact diagonalisation. Data in blue/cyan corresponds to the extended regime where $\deltat$ is finite as $L\to\infty$; whereas the localised regime data, shown in red/orange, shows that $\deltat$ decays to zero as $L^{-1}$ (indicated by the red dashed lines). For the AAH model, we also show  data for the critical point, where $\deltat$ again decays to zero but with an anomalous power, $L^{-0.25}$, shown by the black dashed line. Statistics were accumulated over 5000 values of $\phi \in [0,2\pi)$. }
\label{fig:Ldepend}
\end{figure}

Fig.~\ref{fig:colourmap} shows the resultant $\deltat(\omega)$ and $\yt(\omega)$ as a colour-map in the ($V$-$\omega$)-plane for all three models described in Sec.~\ref{sec:models}. The top row shows $\deltat(\omega)$, a finite value of which indicates an extended phase. On the `other' side of exactly known mobility edges (shown by black dashed lines), $\deltat$ drops to a vanishingly small value, indicating a localised phase. This is consistent with the behaviour of $\yt(\omega)$, shown in the bottom row. In the localised phase, $\yt(\omega)$  is finite and diverges at the mobility edge, indicating a transition to an extended phase. While the results in Fig.~\ref{fig:colourmap} were for a single system size, in Fig.~\ref{fig:Ldepend} we show the system-size dependence of $\deltat(\omega)$ at exemplary $(V,\omega)$ points in the extended and localised phases, for each model. In the extended phase $\deltat(\omega)$ saturates with increasing $L$ to a finite value, whereas in the localised phase it decays to zero as $L^{-1}$. Consistently, $\yt^{-1}(\omega)$ decays to zero in the former and is $L$-independent in the latter as $\yt=\deltat/\eta \propto L\deltat$ $ \sim \mathcal{O}(1)$. The numerical results thus confirm the validity and applicability of $\deltat(\omega)$ and $\yt(\omega)$ as diagnostics for the localisation transitions and mobility edges in the quasiperiodic models considered.

\subsection{Continued fraction \label{sec:cf}}
We now discuss in more detail the continued fraction in Eq.~\eqref{eq:Scontf}. Using $\Delta_0 = -\mathrm{Im}[S_0]$, this
yields a series for $\Delta_0$ as
\eq{
    \Delta_0^{\pd} = \sum_{n=1}^\infty \Delta_0^{\pd}(n)\quad\mathrm{with}\quad \Delta_0^{\pd}(n)\equiv\eta\prod_{k=1}^n\frac{J^2}{\Omega_k^2}
    \label{eq:Delta0-series}
}
and $\Omega_k^2 = (\omega-V\epsilon_k-X_k^{(k-1)})^2 + (\eta+\Delta_k^{(k-1)})^2$. The apparent simplicity of this series is deceptive, as $\Omega_k$ explicitly depends on $X_k^{(k-1)}$ and $\Delta_k^{(k-1)}$ which have their own continued-fraction representations. Physically, $\Delta_0(n)$  is the contribution to $\Delta_0$ of a process where the particle goes out to site $n$ from the root site 0 and retraces its path back to the root~\cite{roy2020localisation}. Such path contributions are also present in the forward-scattering approximation~\cite{pietracaprina2016forward}, albeit in an un-renormalised fashion. Here by contrast, the presence of $X_k^{(k-1)}$'s and $\Delta_k^{(k-1)}$'s in the $\Omega_k$ denominators  reflects that the theory is 
fully (and exactly) renormalised.

In analysing these continued fractions, it proves convenient to introduce a notation
\eq{
    S_0^{[\ell]} = \cfrac{J^2}{\omega^+-V\epsilon_1-\cfrac{J^2}{\omega^+-V\epsilon_2-\cfrac{J^2}{\ddots \cfrac{J^2}{\omega^+-V\epsilon_\ell - S_\ell}  }}}\,,
\label{eq:Scontf-l}
}
where the superscript $[\ell]$ denotes that we have merely iterated the continued fraction to level $\ell$, 
but not truncated it. The $S_\ell$ on the right-hand side has its own continued fraction representation, such that Eq.~\eqref{eq:Scontf-l} is formally exact and continues {\textit{ad infinitum}} for a thermodynamically large system. 
By a truncation at order $\ell$, we shall mean setting $S_\ell=0$, which is physically equivalent to  truncating the half-chain at site $\ell$. 

The localised phase offers significant simplifications in the structure of the continued fractions. Since $\Delta_0\propto\eta$ in this phase and $\eta\to 0^+$, the series for $y_0$ reads
\eq{
    y_0^{\pd} = \sum_{n=1}^\infty y_0^{\pd}(n)\quad\mathrm{with}\quad y_0^{\pd}(n)\equiv\prod_{k=1}^n\frac{J^2}{\Omega_k^2}
    \label{eq:y0series}
}
where $\Omega_k^2 = (\omega-V\epsilon_k-X_k^{(k-1)})^2$ now. Crucially, the coresponding continued fraction for $X_k^{(k-1)}$ is
\eq{
    X_k^{(k-1)} = \cfrac{J^2}{\omega-V\epsilon_{k+1}-\cfrac{J^2}{\omega-V\epsilon_{k+2}-\cfrac{J^2}{\ddots}}},
}
and no longer depends on the $\Delta_k$'s. Hence, the $\Omega_k$'s have a closed set of recursion relations
\eq{
    \Omega_k^{\pd} = \omega-V\epsilon_k^{\pd}-J^2/\Omega_{k+1}^{\pd}.
    \label{eq:Omega-recursive}
}
Truncation of the series Eq.~\eqref{eq:y0series} at level $\ell$ by setting $S_l=0$ corresponds to a boundary condition of the recursion relation Eq.~\eqref{eq:Omega-recursive} as $\Omega_\ell=\omega-V\epsilon_\ell$. Note from Eq.~\eqref{eq:y0series} and Eq.~\eqref{eq:Omega-recursive} that the framework governing $y_0(\omega)$ in the localised phase does not depend on $\eta$, and hence the convergence properties of the series can be analysed directly in the thermodynamic limit, not just formally but also in practice.


\section{Convergence of continued fraction \label{sec:convcf}}

\begin{figure}
\includegraphics[width=\linewidth]{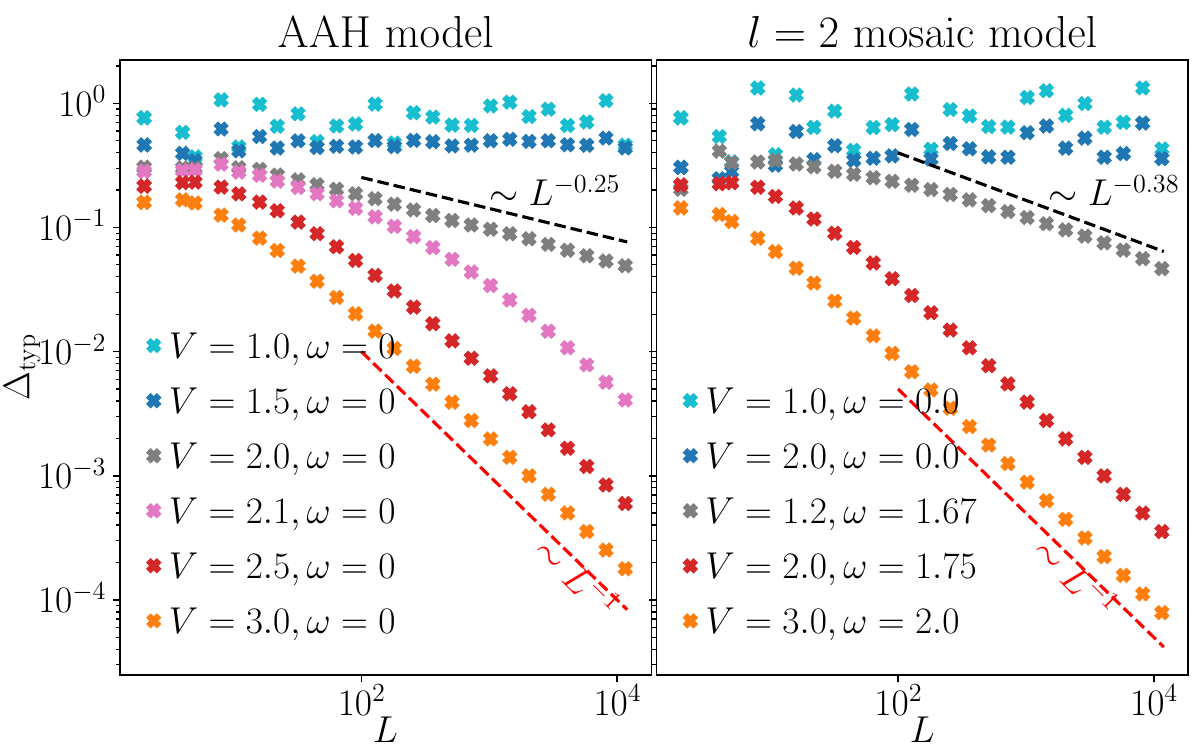}
\caption{Results for $\deltat$ obtained from the exact continued fraction for $S_0^{[\ell]}$ in Eq.~\eqref{eq:Scontf-l} with $\ell=L-1$ for a system with $L$ sites, for the AAH (left) and $l=2$ mosaic (right) models. Data in blue/cyan and pink/red/orange correspond respectively to extended and localised regimes, while data in grey corresponds to a critical point/mobility edge. In the extended regime, $\deltat$ is finite as $L\to\infty$ whereas in the localised regime it decays to zero as $L^{-1}$. At the critical point/mobility edge, the scaling of $\deltat$ with $L$ is anomalous, reflecting the fractal nature of the eigenstates therein. Statistics for all data were obtained over 5000 values of $\phi\in[0,2\pi)$.}
\label{fig:Delta-typ-CF}
\end{figure}

Results for $\deltat$ and $\yt$ are now presented, obtained via the continued fractions discussed in Sec.~\ref{sec:cf}.
We begin with $\deltat$ obtained from Eq.~\eqref{eq:Scontf-l}. For a finite system of size $L$ (with sites $i=0,1,\cdots,L-1$), 
$S_0^{[L-1]}$ is exact, and our approach is to study $\deltat$ thereby obtained as a function of $L$ for different points in the $(V,\omega)$-plane. Representative results are shown in Fig.~\ref{fig:Delta-typ-CF} for the AAH and $l=2$ mosaic model 
(results for the $\beta$-model are omitted for brevity, as they are qualitatively similar to those for AAH). In the extended regime, $\deltat$ as expected is finite in the limit of $L\to\infty$. In the localised regime on the other hand, it decays to zero $\propto L^{-1}$. It is in fact important  that the $L^{-1}$ decay is universal throughout the localised regime, since this in turn implies that $\yt=\deltat/\eta\propto L\deltat$ is finite. 

At the critical point/mobility edges shown in grey in Fig.~\ref{fig:Delta-typ-CF}, $\deltat$ again vanishes as $L\to \infty$
but with a power that is anomalous, $\deltat\sim L^{-\alpha}$ with $\alpha<1$. This we attribute to the (multi)fractal nature of the critical eigenstates. Fractal eigenstates exhibit behaviour  intermediate between perfectly extended and exponentially localised states, as indeed is reflected in the anomalous scaling of $\deltat$ with $L$ at the critical point. As a consistency check, we mention that the anomalous exponents shown in Fig.~\ref{fig:Delta-typ-CF} are identical (within fitting errors) to those obtained from the scaling of $\deltat$ obtained via exact diagonalisation (see e.g.\ Fig.~\ref{fig:Ldepend} for the AAH model). 
\new{The scaling of $\deltat$ with $L$ in the two phases and at criticality 
might tempt one to make a connection to the inverse participation ratios (IPR). IPRs are a common diagnostic for localisation, as they scale as $L^{-1}$ and $L^0$ in the extended and localised phases respectively. Indeed, the results above suggest that $\deltat\sim (L\times \mathrm{IPR})^{-1}$ in either phase. However, such a relation does not hold generally. 
The absence of a general connection between $\deltat$ and the IPRs is in fact evident from their scalings at the critical point. For instance, for the AAH model at criticality, $\deltat\sim L^{-0.25}$ (Fig.~\ref{fig:Delta-typ-CF}), whereas it is readily shown numerically that $\text{IPR}\sim L^{-0.5}$.
}

\begin{figure}[!t]
\includegraphics[width=\linewidth]{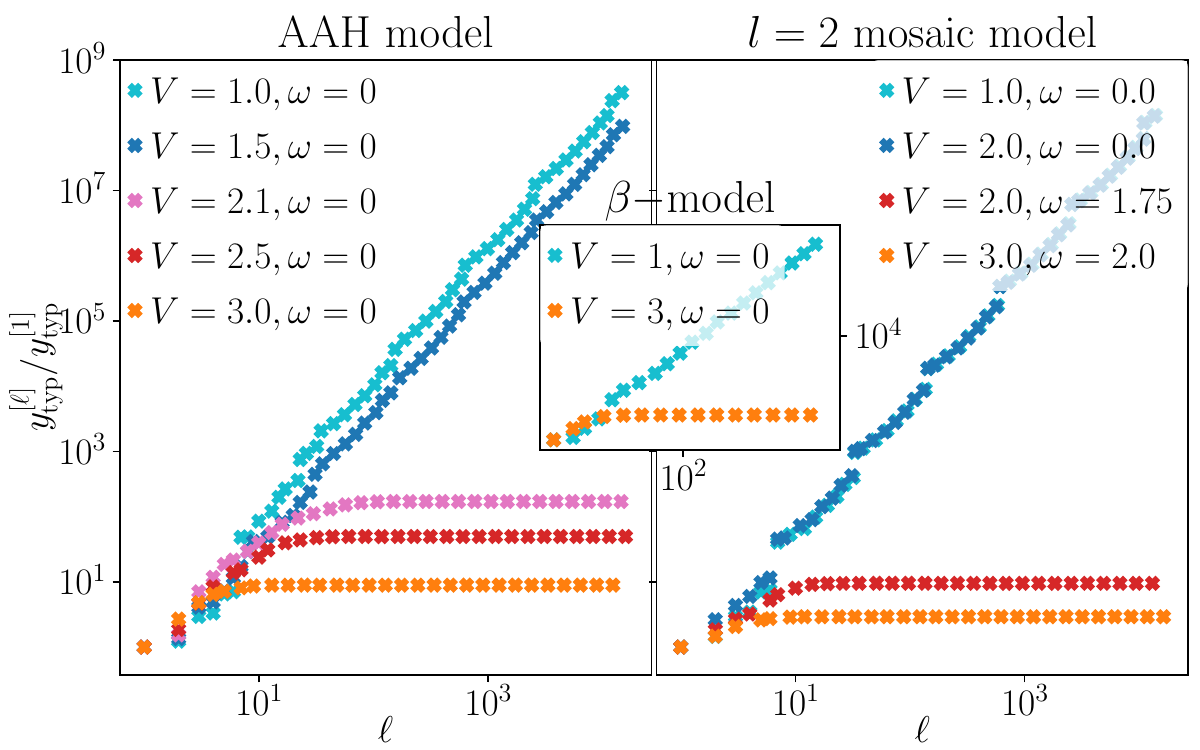}
\caption{Results for $\yt^{[\ell]}=\sum_{n=1}^{\ell} y_{0}(n)$ \emph{vs} $\ell$, obtained for the series in Eq.~\eqref{eq:y0series}. Different panels correspond to the three models, as indicated. In the localised regime $\yt^{[\ell]}$ saturates to a finite value as $\ell\to\infty$. In the extended regime, $\yt^{[\ell]}$ thus computed diverges, 
indicating both the breakdown of localisation and the validity of Eq.~\eqref{eq:y0series}.}
\label{fig:y0-CF}
\end{figure}

We turn next to the continued fraction for $y_0$, Eq.~\eqref{eq:y0series}, which is valid  specifically  in the localised phase. As noted in Sec.\ \ref{sec:cf} in formulating Eqs.~\eqref{eq:y0series} and \eqref{eq:Omega-recursive}, the computation of $y_0$ can be interpreted as one directly in the thermodynamic limit. One therefore studies the convergence of the series with the truncation level $\ell$, writing (see Eq.~\eqref{eq:y0series}) $y_{0}^{[\ell]}=\sum_{n=1}^{\ell} y_{0}(n)$ and studying the convergence of the series with increasing $\ell$. Saturation of the series to a finite value as $\ell\to\infty$ indicates a localised regime, whereas in an extended regime the framework itself breaks down, leading to a divergent $y_{0}^{[\ell]}$
with increasing $\ell$.  The results of Fig.~\ref{fig:y0-CF}, which show the $\ell$-dependence of  the resultant typical value 
$y_{\mathrm{typ}}^{[\ell]}$, indeed confirm this behaviour.

In the localised regime, while $\yt$ saturates to a finite value as $\ell\to\infty$, Fig.~\ref{fig:y0-CF} shows that
the saturation value, as well as the length-scale at which the saturation occurs, grows with proximity to the critical point/mobility edge. This in turn reflects physically the fact that the typical localisation length in the $(V,\omega)$-plane increases as one moves closer to a transition. In order to understand this in the simplest fashion, consider an eigenstate 
$|\psi_{n}\rangle$ of some energy $\w$, which is localised on the root site with a localisation length $\xi$.
The (normalised) wavefunction densities $\vert\psi(r)\vert^2 =|\langle r|\psi_{n}\rangle|^{2}$ for such a state can then be written as $\vert\psi(r)\vert^2 = (1-e^{-1/\xi}) e^{-r/\xi}$.  At the same time, it is straightforward to show~\cite{ThoulessReview1974} that $\vert\psi(0)\vert^2 = (1+y_0(\omega))^{-1}$, which leads to a relation between $y_0(\omega)$ and 
$\xi$, viz.\  $y_0(\omega) = (1-e^{-1/\xi})^{-1}-1$ $\overset{\xi \gg 1}{\sim} \xi$; showing explicitly that an increasing typical $\xi$ implies an increasing $\yt$.  Moreover near a critical point, where $\xi \gg 1$, $y_0(\omega)\sim \xi$ shows that the typical localisation length and $\yt$ each diverge with the same critical exponent  ($\nu =1$), 
on approaching a transition from the localised side.

The same physical picture can be used to understand that $\yt^{[\ell]}$ saturates when $\ell$ exceeds a length-scale set by
the typical localisation length. As discussed  previously, $y_0(n)$ in Eq.~\eqref{eq:y0series} is the contribution to $y_0$ of processes where the particle goes out to site  $n$ and retraces its path on the chain back to the root. However, since the wavefunction density decays exponentially with $r$ with a length-scale $\xi$, then naturally for all $n\gg \xi$ the contributions are negligibly small. The sum in Eq.~\eqref{eq:y0series} therefore saturates for $n\gtrsim\xi$ which, crucially, is independent of $L$. 
We will return to this briefly in the next section.

Finally  here, as explained in Sec.~\ref{sec:cf}, the series Eq.~\eqref{eq:y0series} is not applicable in the extended regime.
Nevertheless, studying the series in an extended regime indeed signals the breakdown of localisation~\cite{roy2020localisation}, via the fact that $\yt^{[\ell]}$ diverges as $\ell\to\infty$.


\section{Self-consistent theory \label{sec:sct}}
In the previous section, truncation of the continued fraction was enacted by setting the terminal $S_\ell$ to zero, and analysing its convergence as a function of truncation order $\ell$. An alternative approach
is to assign a typical value to the terminal self-energy $S_\ell\rightarrow -i\deltat$, and analyse the series self-consistently. This amounts to having a continued fraction for $\Delta_0$ which depends on all the quasiperiodic site energies up to site $\ell$, together with $\deltat$ on site $\ell$. Compiling statistics over $\phi$, one thus obtains a distribution of $\Delta_0$, $P_\Delta(\Delta_0,\deltat)$, which depends parametrically on $\deltat$. Self-consistency is then imposed by requiring that $\deltat$ obtained from this distribution is equal to the parametric $\deltat$,
\eq{
    \log\deltat = \int d\Delta_0^{\pd}~P_\Delta^{\pd}(\Delta_0^{\pd},\deltat)\log\Delta_0^{\pd}.
}
This constitutes a self-consistent analysis at order $\ell$. One can analogously construct a similar self-consistent framework for $\yt$. In Ref.~\cite{duthie2020selfconsistent} we analysed the self-consistent theory at leading order $(\ell=1)$;
resulting, for the models specified in Sec.~\ref{sec:models}, in analytical results for mobility edges and phase diagrams
that are in very good agreement with previous results. 
Here, by contrast, our focus will be on higher-order theories, and in particular how their convergence is dictated by the convergence of the underlying CF.

First, however, we point out briefly that self-consistency imposed separately in the extended and localised phases breaks down at the same point in the $(V,\omega)$-plane, demonstrating a consistent critical point/mobility edge within the theory.
In terms of the series in Eqs.~\eqref{eq:Delta0-series} and \eqref{eq:y0series}, one obtains respectively
\eq{
    \begin{split}
    \Delta_0^{\pd} &= \Delta_0^{[\ell-1]} + \Delta_0^{\pd}(\ell)(1+\deltat/\eta),\\
    y_0^{\pd} &= y_0^{[\ell-1]} + y_0^{\pd}(\ell)(1+\yt).
    \end{split}
    \label{eq:sc-series}
}
Near a localisation critical point, where a typical $\Delta_{0} \to 0$, one can  asymptotically replace
$\Omega_k\to (\omega-V\epsilon_k-X_k^{(k-1)})$, which reduces the two equations in Eq.~\eqref{eq:sc-series} to the same equation. 
The same critical point thus arises as the transition is approached from either phase. This is also reflected in the fact that the distribution $P_\Delta(\Delta)$ in the extended regime smoothly evolves to $P_y(y)$ in the localised regime (after  a trivial rescaling by $\eta$).  At leading order ($\ell =1$) it can in fact be shown explicitly~\cite{duthie2020selfconsistent} that the self-consistent equations in  the extended and localised regimes are respectively
\eq{
    \begin{split}
    &\langle\langle \ln[(\omega-V\epsilon_1^{\pd})^2+\deltat^2]\rangle\rangle-\ln J^2=0\,,\\
    &\ln(1+\yt^{-1}) = \langle\langle\ln(\omega-V\epsilon_1^{\pd})^2\rangle\rangle-\ln J^2\, ,
    \end{split}
    \label{eq:leading-sc}
}
 where $\braket{\braket{\cdot}}$ denotes an average over $\phi$ and kinds of end-sites. Requiring  $\deltat,\yt>0$ (as 
$\Delta_{0}$ is physically a rate), one obtains an expression for the mobility edge as
\eq{
    \langle\langle \ln[(\omega_\mathrm{ME}^{\pd}-V\epsilon_1^{\pd})^2]\rangle\rangle-\ln J^2=0\,.
    \label{eq:ME-general}
}
While such a simple and tractable expression for the mobility edge is not  yielded by higher-order self-consistent theories, they can be readily implemented numerically; and we will be particularly interested in showing that they converge to the same results as the analysis using the continued fractions.

Let us now establish the conceptual connection between the two approaches. In the extended phase, the series for $\Delta_0$ typically converges to a finite value. In order for this to happen, successive terms $\Delta_0(n)$ in the series  must decay with 
increasing $n$ sufficiently rapidly, such that $\Delta_0(n)\to 0$ as $n\to\infty$. The effect on $\Delta_0$ of multiplying $(1+\deltat/\eta)$ by $\Delta_0(\ell)$ in Eq.~\eqref{eq:sc-series}, is therefore negligible provided $\ell$ is large enough that $\Delta_0(\ell)/\eta \to 0$. The self-consistent solution of $\deltat$ is then governed entirely by the series, and one naturally expects the two approaches to yield identical results. In the localised regime, $\deltat\to 0$. Since the localisation length, $\xi$, is finite, it is obvious that $\Delta_0$ is insensitive to either cutting off the chain at a site $\ell$ or endowing $S_\ell$ with an imaginary part (provided $\ell\gg \xi$), again leading to an equivalence of the two approaches.

\begin{figure}
\includegraphics[width=\linewidth]{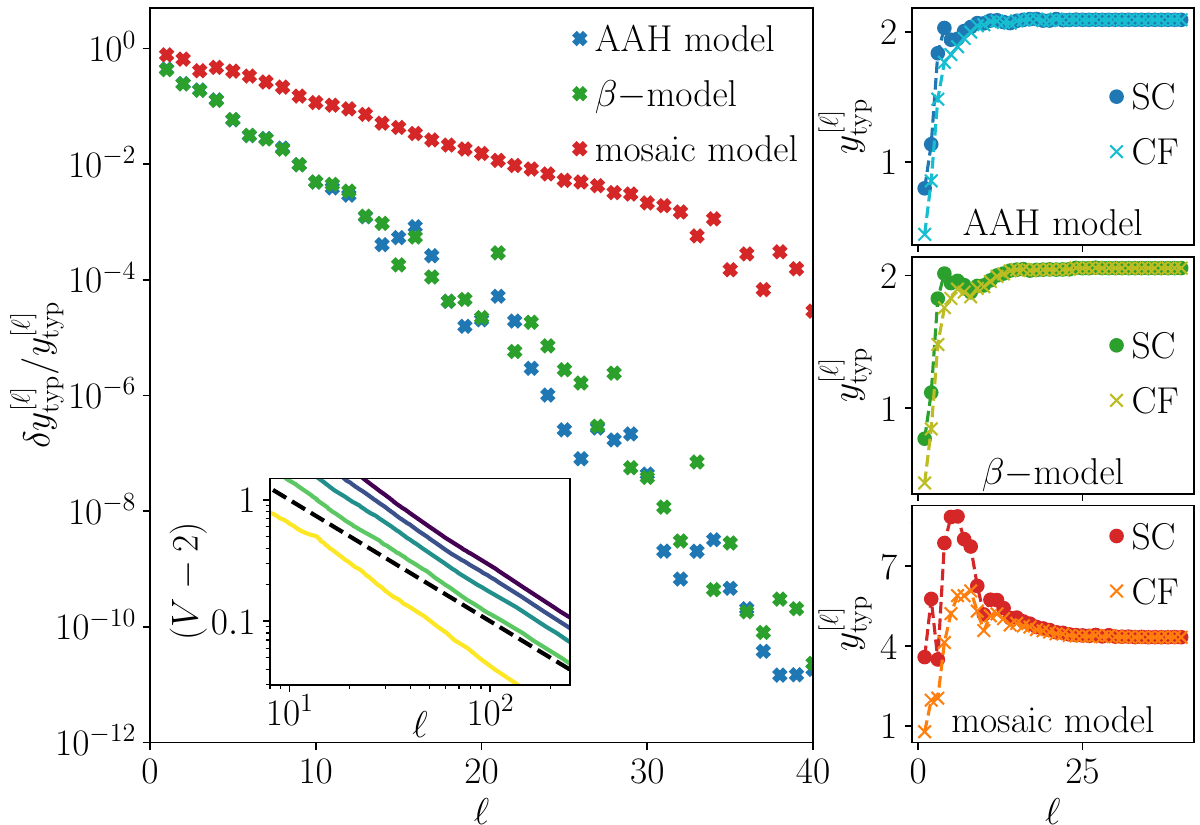}
\caption{\new{Convergence of $\yt$, obtained via the self-consistent theory (SC) and continued fraction (CF), to the same result with 
increasing order of truncation ($\ell$) in the localised regime. Left panel shows that the relative difference, 
$\delta \yt^{[\ell]}/\yt^{[\ell]}$, where $\delta \yt^{[\ell]} =  \yt^{[\ell],\mathrm{SC}} -  \yt^{[\ell],\mathrm{CF}}$, between the SC and CF decays exponentially with $\ell$, while the right panels show the raw data for $\yt^{[\ell]}$ for  each model. Results are shown for $V=3$ and 
$\omega=0$ for the AAH and $\beta=0.2$ models, and for $V=1.7$ and $\omega=1.5$ for the mosaic model. The inset in the left panel shows the average $\ell$ required to achieve convergence of the CF to within a given threshold. The differently coloured lines correspond to $\delta\yt^{[\ell],\text{CF}} = 10^{-2}$, $10^{-4}$, $10^{-6}$, $10^{-8}$ and $10^{-10}$ from yellow to purple, where  $\delta\yt^{[\ell],\text{CF}} = \yt^{[\ell+1],\text{CF}}-\yt^{[\ell],\text{CF}}$. The black dashed line denotes $(V-V_c)^{-1}$.
}}
\label{fig:sc-cf-conv}
\end{figure}

The connection between the two appoaches can likewise be understood in terms of $\yt$, which is the quantity of interest in the localised regime. There, the series for $y_{0}$  in Eq.~\eqref{eq:y0series} saturates to a finite value, implying that $y_0(n)$ decays sufficiently rapidly with increasing $n$. Analogous to the argument given above, capping $y_0(\ell)$ with $(1+\yt)$ is thus immaterial for $y_0$ in Eq.~\eqref{eq:sc-series} provided $\ell\gg\xi$. The two approaches are therefore bound to give the same result for $\yt$.  We confirm this as shown in Fig.~\ref{fig:sc-cf-conv}, by comparing the $\ell$-dependence of the  numerically evaluated $\yt^{[\ell]}$, to the $\yt$ arising from the self-consistent theory at order $\ell$. Defining
$\delta \yt^{[\ell]} =  \yt^{[\ell],\mathrm{SC}} -  \yt^{[\ell],\mathrm{CF}}$ as the difference\footnote{
$\delta \yt^{[\ell]}=y_0(\ell)\yt $ is necessarily positive, since $\yt,y_0(\ell)>0$.} between the results from the self-consistent (SC) and continued fraction (CF) approaches, we find that 
\new{the relative error $\delta \yt^{[\ell]}/\yt^{[\ell]}$} decays to zero exponentially with $\ell$ (Fig.~\ref{fig:sc-cf-conv}, left panel), showing that the two approaches converge to the same solution.
This is also directly evident in the right panels of Fig.~\ref{fig:sc-cf-conv}, where the $\ell$-dependence of
$y_{\mathrm{typ}}^{[\ell]}$ is shown for each approach, and all three models considered. 
\new{
    As mentioned above, for $\ell\gg\xi$, the two approaches converge to the same (exact) result for $\yt$. However, on approaching the critical point/mobility edge from the localised side, the localisation length grows unboundedly. Hence one expects the $\ell$ required for the CF to converge to increase as one gets closer to the critical point. This is indeed the case, as shown in the inset of Fig.~\ref{fig:sc-cf-conv}, which shows the average value of $\ell$ required as a function of $V$ to achieve a specific threshold for the error in $\yt$. We find that for sufficiently small thresholds, the required $\ell$ diverges as $(V-V_c)^{-1}$ which is exactly the form with which the localisation length diverges in this model~\cite{thouless1983bandwidths}.
}

\section{Self-energy Distributions and spectra from higher-order theories  \label{sec:dists}}

\begin{figure}
\includegraphics[width=\linewidth]{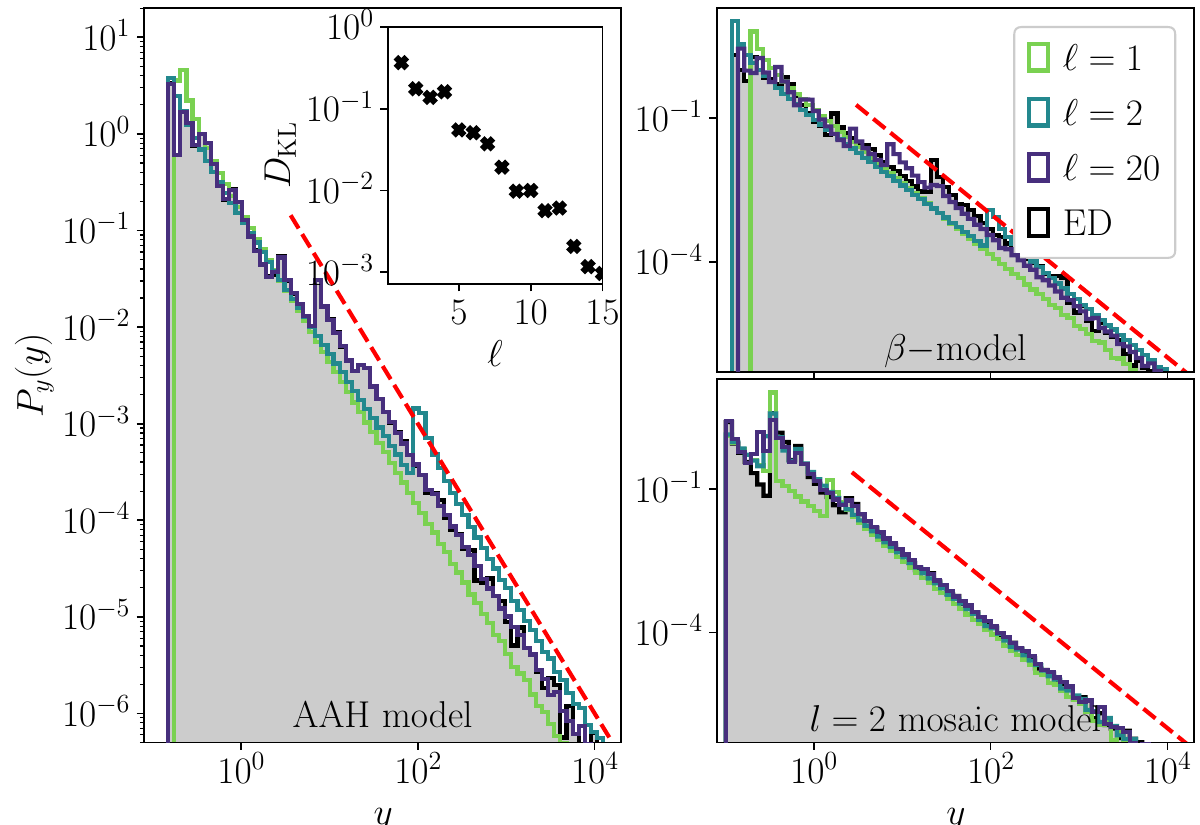}
\caption{Distributions $P_y(y)$  \emph{vs} $y$ obtained from the self-consistent (SC) theory truncated at order $\ell$
(shown for $\ell =1,2,20$),  and compared to the exact result obtained from ED for $L=5000$ (shaded grey with black edging).  Note that the $\ell =20$ SC results and those from ED are barely distinguishable. With increasing $\ell$ the additional structures on top of the $\propto y^{-3/2}$ L\'evy tail (shown by the red dashed line) are recovered. The inset in the left panel shows the Kullback–Leibler divergence between the distribution obtained from the SC method and  that obtained from ED. It decays rapidly with $\ell$, indicating rapid convergence of the distributions to the exact result. Results are shown for $V=3$ for all three models, with 
$\omega=0$ for the AAH and $\beta=0.2$ models and $\omega=2$ for the mosaic model.}
\label{fig:y-dists}
\end{figure}

Since $\deltat$ and $\yt$ are probabilistic order parameters, their distributions are naturally important objects to study. These can be obtained analytically at leading order, revealing a seemingly universal characteristic L\'evy-tailed distribution ($\propto y^{-3/2}$ for $y\gg 1$)~\cite{duthie2020selfconsistent}. However, such an approach fails to capture additional structures 
residing on top of the L\'evy tail, originating from and reflecting the quasiperiodic nature of the potential. In fact, quasiperiodic systems of the kind considered here also have important fine  structures in their eigenvalue spectra  (or total DoS), in  the  form  of  hierarchical gaps. In this section, we show going to higher orders in the theory increasingly captures these structures.

\subsection{Self-energy distributions}

We will focus mainly on the distributions $P_y(y)$, owing to their apparent universality for disordered systems (with both uncorrelated~\cite{abou-chacra1973self} and correlated~\cite{roy2020localisation} disorder) as well as 
quasiperiodicity.
Within the self-consistent theory these distributions follow directly, because once the self-consistent $\yt$ is obtained the distribution $P_y(y,\yt)$ is specified.

Comparison of $P_y(y)$ obtained from exact diagonalisation (ED) and from the leading order theory ($\ell=1$), as shown in Fig.~\ref{fig:y-dists}, shows that the latter does not capture the evident finer structure in the distributions (on top of the 
background L\'evy tail). These can be attributed to the deterministic and infinite-range correlated nature of the quasiperiodic potential, as they are not washed away via $\phi$-averaging. We thus expect that higher-order theories, which explicitly take into account the potentials on all sites up to the truncation order, will reveal them.

Fig.~\ref{fig:y-dists} 
accordingly
shows the distributions $P_y(y)$ obtained from the self-consistent theory in the localised phase, for different values of truncation order $\ell$. While the $y^{-3/2}$ tail is universally present at all orders and for all three models, with increasing $\ell$ finer structures are indeed seen to emerge in the distributions, which match very well
with the exact result obtained from ED
(and which we have checked are converged  with respect to $\phi$-averaging).
This confirms that these structures are manifestations of the highly-correlated 
quasiperiodic potential, and as such they are captured at higher orders of the theory.

\begin{figure}
\includegraphics[width=\linewidth]{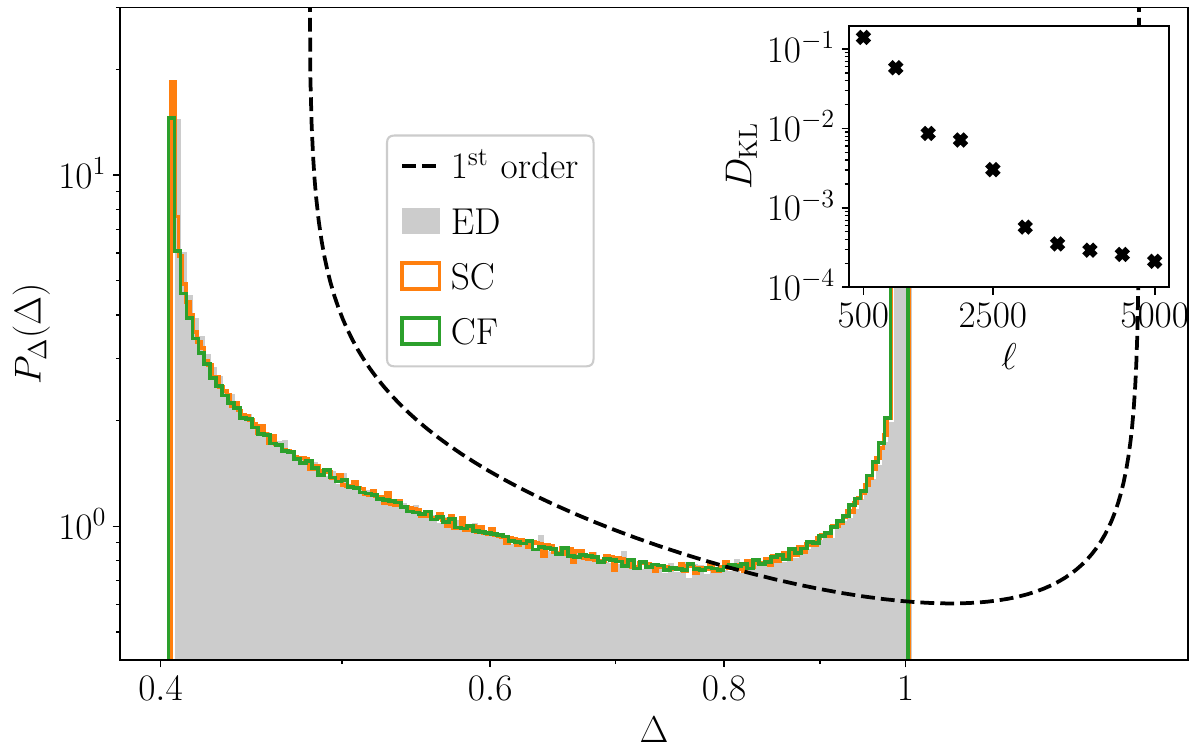}
\caption{
Distributions $P_\Delta(\Delta)$ for the AAH model in the extended phase (shown for $V=1$ and $\omega=0$),
obtained from the continued fraction (CF) and self-consistent (SC) treatment at order $\ell=5000$, together with results obtained from ED for $L=5000$. The three are barely distinguishable. Inset panel shows the Kullback-Leibler divergence 
$D_\mathrm{KL}$ between SC distributions at various orders $\ell$ and the ED distribution. $D_\mathrm{KL}$ decays with increasing $\ell$,  indicating convergence of the self-consistent distributions to the ED result. The analytic leading-order 
($\ell =1$) SC result is also shown (black dashed line), and bar an overall shift captures the distribution rather well.
}
\label{fig:deloc_CF_dists}
\end{figure}

To quantify the convergence of the distributions obtained to those  obtained exactly from ED, we compute the Kullback–Leibler divergence~\cite{kullback1951information}
\eq{
    D_{\mathrm{KL}}^{\pd}(\ell) = \int dy~P^\mathrm{ED}_y(y)\log\left(\frac{P^\mathrm{ED}_y(y)}{P^{[\ell]}_y(y)}\right),
    \label{eq:dkl}
}
where $P^{[\ell]}_y$ is the distribution obtained from the self-consistent theory at order $\ell$. As shown in the inset in 
Fig.~\ref{fig:y-dists}, $D_{\mathrm{KL}}(\ell)$ decays rapidly to zero with $\ell$ showing the rapid convergence of the distributions. We add that the distributions obtained from the CF approach likewise converge to the ED results at large $\ell$.

Finally, to exemplify results in the extended regime, Fig.~\ref{fig:deloc_CF_dists} shows representative 
distributions of $P_{\Delta}(\Delta)$ for the AAH model.  Obtained as specified in the caption from the truncated continued fraction method, the self-consistent approach, and ED calculations, the three are seen to be barely distinguishable from each other. The figure also shows the analytical result arising~\cite{duthie2020selfconsistent} from the leading order ($\ell$$=$$1$) self-consistent theory  (with characteristic square-root singularities evident at the hard edges of the distribution).  
Excepting an overall shift, this is seen to capture the full distribution rather well.

\subsection{Hierarchical gaps in spectra}

\begin{figure}
\includegraphics[width=\linewidth]{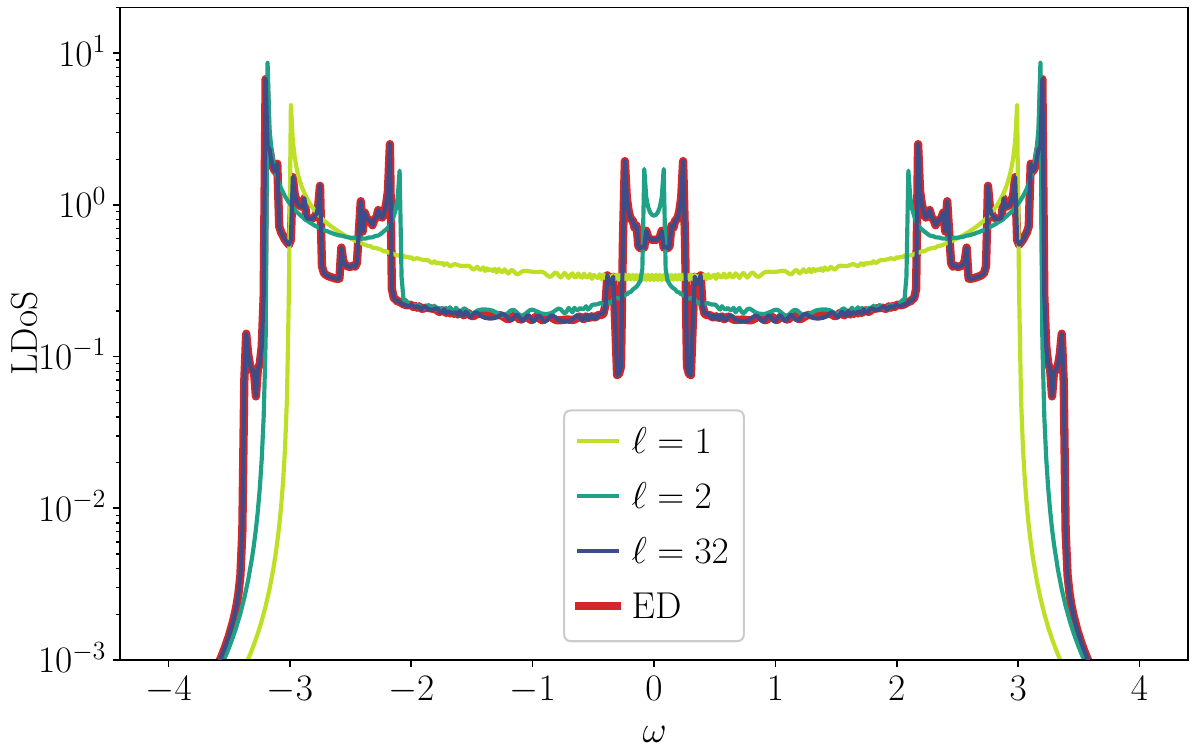}
\caption{The average LDoS for the  root site (shown for the AAH model, with $V=3$),  showing hierarchical structure in the mean local spectrum. While these features are not captured by the continued fraction at low orders ($\ell$), the higher-order theories capture them well, and are well converged to the ED result. The latter is shown here for $L=1024$. }
\label{fig:mean-ldos}
\end{figure}

While the total eigenvalue spectrum, or DoS, does not by itself reveal the localised or extended nature of the states,
the presence of hierarchical gaps in the spectra of quasiperiodic systems is 
an important intrinsic characteristic of them, and is directly  evident in the ED results shown in Fig.~\ref{fig:colourmap}. The theory presented here can also be used to probe such structure. We give an illustrative
example in Fig.~\ref{fig:mean-ldos}.  For the AAH model, this shows the $\phi$-averaged local DoS for the root site
(which is proportional to the imaginary part of the local self-energy); obtained from the continued-fraction approach
with increasing order $\ell$, and compared to ED results.  At the lowest-order level ($\ell =1$) the mean LDoS is smooth, and devoid of the peaks obtained by ED which reflect the actual energy bands. On increasing $\ell$ however, the peaks begin to appear, and by a modest value of $\ell=32$ the mean LDoS obtained from the continued fraction is well converged to the numerically exact result.

\section{Discussion \label{sec:discussion}}

In summary, we have presented a theory for localisation in quasiperiodic chains with nearest-neighbour hoppings, 
 based upon the continued-fraction representation of the local propagators, and in particular the end-site propagators for
a semi-infinite half-chain. Our focus was on the imaginary part, $\Delta(\w)$, of the associated self-energy, which likewise admits a continued-fraction representation and which acts as a  probabilistic order parameter. In a regime of extended states, we showed that the typical value of the self-energy distribution, $\deltat(\w)$, converges to a finite value; while in a localised regime it decays to zero as the level of the continued fraction is increased. Analogously, the continued fraction for the `complementary' order parameter $\yt(\w)=\deltat(\w)/\eta$ was shown  to converge to a finite value in the localised regime but diverges in the extended regime. Together, they can be used to map out the localisation phase diagram in the space of energy and Hamiltonian parameters, thus giving access to the critical points/mobility edges. Interestingly, it was also found that at the critical points, $\deltat(\w)$ has an anomalous scaling with the system size and decays to zero as $L^{-\alpha}$ with $\alpha<1$, which we attribute to the fractal nature of the critical eigenstates. The continued-fraction method was moreover shown to be 
intimately connected to higher-order self-consistent theories;  such that the two approaches are asymptotically equivalent,
and give quantitative agreement with results arising from exact diagonalisation. Finally, we showed how going to higher orders in the theory reveals finer structures in the distributions of $\Delta$ and $y$,   reflecting the deterministic and highly correlated nature of the quasiperiodic potential. These substructures appear on top of the universal features such as a 
L\'evy-tailed distribution of $y$ in the localised regime. By the same token, the higher-order theories also reveal the hierarchical gaps in the disorder-averaged spectra of these systems.

In the continued-fraction approach, the series was truncated by setting the terminal self-energy $S_\ell \to0$. In the 
self-consistent approach on the other hand, $S_\ell\to -i\deltat$ was set to a typical value. In both approaches, the regulator played a central role. It is worth noting an approach similar in spirit that has been employed for disordered 
systems~\cite{miller1994weak,monthus2008anderson,garciamata2017scaling}, in which one avoids making $\omega$ complex via the regulator, and self-consistency is not imposed. The imaginary part of the self-energy is simply seeded by assigning a finite and constant imaginary terminal self-energy,  $S_\ell\to -i$. Physically, this corresponds to connecting the terminal site to a 
conducting lead. The quantity of interest then becomes the imaginary part $\Delta_0$ that is induced at the root site.
In a localised regime, a finite localisation length implies that the imaginary part in $S_\ell$ does not in effect propagate 
to $S_0$,  such that $\Delta_0\to 0$ for $\ell\gg\xi$; while in an extended regime the effect of coupling to the conducting lead induces a non-vanishing $\Delta_{0}$.

One qualitatively new result found from the higher-order theories is the anomalous scaling of $\deltat$ with $L$ at critical points/mobility edges. Looking to the future, establishing a quantitative connection between this and the anomalous scaling of generalised inverse participation ratios of critical eigenstates, is naturally a question of substantial interest.
\new{Formally, the residues at the poles of the \textit{full} propagator $G_j(\omega)$ (Eq.~\ref{eq:S_j_def}) give the eigenstate amplitudes on site $j$, from which generalised IPRs may be constructed. } \new{However, the structure of the RPS for the full propagator is naturally more complicated than that of the half-chain propagator, and hence we leave this analysis for a future work.} Such an advance would shed light not only on the critical points of the models considered here, but also on other models which are critical throughout, such as Fibonacci 
chains~\cite{kohmoto1983localization,ostlund1983one}.

Finally, a further interesting direction for future work is the case of models with longer-ranged hopping~\cite{biddle2010predicted,gopalakrishnan2017self,monthus2019multifractality,wang2020duality}, or quasiperiodic models in higher-dimensions~\cite{bodyfelt2014flatbands,danieli2015flat,rossignolo2019localisation}; where the general RPS framework for the self-energies continues
to hold, albeit with a substantially different underlying structure.

\begin{acknowledgments}
We thank I. Creed  for helpful discussions. We are also grateful to the EPSRC for support, under Grant No. EP/L015722/1 for the TMCS Centre for Doctoral Training, and Grant No. EP/S020527/1.
\end{acknowledgments}

\bibliography{refs}

\end{document}